\documentclass[12pt,fleqn]{article}
\usepackage{graphicx,cite,amssymb}

\newcommand{\newsection}[1]{\section{#1}\setcounter{equation}{0}}

\def\be{\begin{equation}}
\def\ee{\end{equation}}
\def\bea{\begin{eqnarray}}
\def\eea{\end{eqnarray}}
\def\nnb{\nonumber}
\def\bbuildrel#1_#2^#3{\mathrel{\mathop{\kern 0pt#1}\limits_{#2}^{#3}}}
\def\slash#1{\setbox0=\hbox{$#1$}#1\hskip-\wd0\dimen0=5pt\advance
       \dimen0 by-\ht0\advance\dimen0 by\dp0\lower0.5\dimen0\hbox
         to\wd0{\hss\sl/\/\hss}}

\newcommand{\f}{\frac}

\newcommand{\me}[1]{\langle#1\rangle}
\newcommand{\al}{\alpha_{\mathrm s}}

\newcommand{\ep}{\epsilon}

\begin{document}
\begin{titlepage}
\begin{flushright}
\vspace*{-3cm}
TTK-23-22,~ TTP23-038,~ P3H-23-063,\\
ZU-TH 57/23,~ ALBERTA-THY-6-23,\\
SI-HEP-2023-20,~ MPP-2023-190\\[2cm]
\end{flushright}
\begin{center}
\setlength {\baselineskip}{0.3in} 
{\bf\Large\boldmath The $Q_{1,2}$--$Q_7$ interference contributions to $b \to s \gamma$ at ${\mathcal O}(\al^2)$ 
                    for the physical value of $m_c$}\\[15mm]
\setlength {\baselineskip}{0.2in}
{\large 
M.\ Czaja$^1$,
M.\ Czakon$^2$,
T.\ Huber$^3$,
M.\ Misiak$^1$,
M.\ Niggetiedt$^4$,
A.\ Rehman$^{5,6}$,\\[1mm]
K.\ Sch\"onwald$^7$
and~
M.\ Steinhauser$^8$}\\[5mm]
$^1$~{\it Institute of Theoretical Physics, Faculty of Physics, University of Warsaw,\\
                    02-093 Warsaw, Poland.}\\[3mm]
$^2$~{\it Institut f\"ur Theoretische Teilchenphysik und Kosmologie, RWTH Aachen University,\\
          52056 Aachen, Germany.}\\[3mm]
$^3$~{\it Theoretische Physik 1, Center for Particle Physics Siegen (CPPS), Universit\"at Siegen,\\
          57068 Siegen, Germany.}\\[3mm]
$^4$~{\it Max Planck Institute for Physics, F\"ohringer Ring 6, 80805 M\"unchen, Germany.}\\[3mm]
$^5$~{\it Department of Physics, University of Alberta, Edmonton, AB T6G 2E1, Canada.}\\[3mm] 		    
$^6$~{\it National Centre for Physics, Quaid-i-Azam University Campus, Islamabad 45320, Pakistan.}\\[3mm]
$^7$~{\it Physik Institut, Universit\"at Z\"urich, 8057 Z\"urich, Switzerland.}\\[3mm] 		    
$^8$~{\it Institut f\"ur Theoretische Teilchenphysik, Karlsruhe Institute of Technology (KIT),\\
          76128 Karlsruhe, Germany.}\\[1cm] 
{\bf Abstract}\\[5mm]
\end{center} 
\setlength{\baselineskip}{0.2in} 

The $\bar B\to X_s\gamma$ branching ratio is currently measured with
around $5\%$ accuracy. Further improvement is expected from Belle
II. To match such a precision on the theoretical side, evaluation of
${\mathcal O}(\al^2)$ corrections to the partonic decay $b \to
X_s^{\rm part}\gamma$ are necessary, which includes the $b \to s
\gamma$, $b \to s g\gamma$, $b \to s gg\gamma$, $b \to sq\bar q\gamma$
decay channels. Here, we evaluate the unrenormalized contribution to
$b \to s \gamma$ that stems from the interference of the photonic
dipole operator $Q_7$ and the current-current operators $Q_1$ and
$Q_2$. Our results, obtained in the cut propagator approach at the
4-loop level, agree with those found in parallel by Fael {\it et
al.} who have applied the amplitude approach at the 3-loop
level. Partial results for the same quantities recently determined by
Greub {\it et al.} agree with our findings, too.

\end{titlepage}

\newsection{Introduction \label{sec:intro}}

Rare $B$-meson decays that receive their leading Standard Model (SM)
contributions from one-loop diagrams provide important constraints on
popular Beyond Standard Model (BSM) scenarios. Among them, the
inclusive radiative decay $\bar B\to X_s\gamma$ is of particular
interest. Its isospin- and ${\mathrm CP}$-averaged branching ratio has been
measured by CLEO~\cite{Chen:2001fja}, Belle~\cite{Belle:2009nth,
Saito:2014das}, BABAR~\cite{Aubert:2007my, Lees:2012wg, Lees:2012ym},
and Belle II~\cite{Belle-II:2022hys} for $E_\gamma > E_0$ in the
decaying meson rest frame, with various values of $E_0$, ranging from
$1.7\,{\rm GeV}$ to $2.0\,{\rm GeV}$. The current world average of
these measurements\footnote{
The most recent Belle II result~\cite{Belle-II:2022hys} is not yet included in the average~(\ref{br.exp}).}
extrapolated to $E_0 = 1.6\,{\rm GeV}$
reads~\cite{HeavyFlavorAveragingGroup:2022wzx,
ParticleDataGroup:2022pth}
\be \label{br.exp}
{\mathcal B}(\bar B\to X_s\gamma)^{\rm exp} = (3.49 \pm 0.19) \times 10^{-4}\, .
\ee
An extrapolation in $E_0$ has been applied because measurements are
less precise for lower values of $E_0$ due to a rapidly growing
background --- see, e.g., Fig.~2 in Ref.~\cite{Belle-II:2022hys}. On the
other hand, theoretical estimates of non-perturbative effects are less
precise for higher values of $E_0$ --- see
Refs.~\cite{Bernlochner:2020jlt, Dehnadi:2022prz} for the most recent
analyses of this issue.

At $E_0 = 1.6\,{\rm GeV}$, the $\bar B\to X_s\gamma$ decay rate is
well approximated by the corresponding perturbative $b \to
X_s^p\gamma$ decay rate, where $X_s^p = s,sg,sgg,sq\bar q,\ldots$ are
the partonic final states. The most relevant non-perturbative corrections
to this approximation are suppressed by powers of $(m_B-m_b)/m_b$.
The largest non-perturbative contribution to the overall uncertainty
arises from the so-called resolved photon effects that have been
extensively studied in Refs.~\cite{Benzke:2010js, Gunawardana:2019gep,
Benzke:2020htm, Benzke:2022cbw, Hurth:2023paz}.

As far as the dominant perturbative contributions are concerned, they
need to be evaluated including Next-to-Leading Order (NLO)
electroweak and Next-to-Next-to-Leading Order (NNLO) QCD corrections to
match the experimental accuracy. Some of the important NNLO QCD
(${\mathcal O}(\al^2)$) corrections that depend on the charm quark
mass $m_c$ have been calculated only in the limits
$m_c=0$~\cite{Czakon:2015exa} and $m_c \gg m_b$~\cite{Misiak:2010sk}.
An interpolation between the two limits was then
applied~\cite{Czakon:2015exa}. The resulting SM prediction obtained in
2015~\cite{Misiak:2015xwa} was subsequently updated in
2020~\cite{Misiak:2020vlo} to yield
\be \label{br.sm}
{\mathcal B}(\bar B\to X_s\gamma)^{\rm SM} = (3.40 \pm 0.17) \times 10^{-4}\, ,
\ee
where the overall uncertainty contains $\pm 3\%$ from the
$m_c$-interpolation, $\pm 3\%$ from unknown higher-order effects, and
$\pm 2.5\%$ from the input parameters (combined in quadrature), which
includes the non-perturbative uncertainties. The resolved photon
contributions are treated along the lines of
Ref.~\cite{Gunawardana:2019gep} --- see Ref.~\cite{Misiak:2020vlo} for
details.

While the results in Eqs.~(\ref{br.exp}) and~(\ref{br.sm}) are in
perfect agreement, further improvement on both the experimental and
theoretical sides are expected. The ultimate Belle II luminosity will
allow for high-statistics measurements using the hadronic tag
method for the recoiling $B$ meson, which efficiently suppresses the
non-$B\bar B$ backgrounds~\cite{Kou:2018nap} and makes the
determination of $E_\gamma$ in the decaying $B$-meson rest frame
possible on an event-by-event basis~\cite{Aubert:2007my,
Belle-II:2022hys}. As far as the perturbative calculations are
concerned, most effort is being devoted to eliminating the
$m_c$-interpolation at ${\mathcal O}(\al^2)$ by evaluation of the
corresponding corrections at the physical value of $m_c$.

In this paper, we present results of our calculation of the
unrenormalized $m_c$-dependent NNLO QCD corrections to the $b \to
s\gamma$ decay rate at the physical value of $m_c$. They need to be
supplemented in the future with the corresponding bremsstrahlung
contributions (with $X_s^p = sg,sgg,sq\bar q$), for which our
calculations are advanced but not yet finished. However, the observed
agreement with the parallel calculation of Ref.~\cite{Fael:2023gau},
as well as the published partial results of Ref.~\cite{Greub:2023msv}
makes us confident that the two-body-final-state contributions in our
calculation can be treated as cross-checked, and are ready for
publication.

The article is organized as follows. In the next section, we provide
the necessary definitions to specify the corrections we actually
calculate. Next, our method for the evaluation of 4-loop propagator
diagrams with unitarity cuts is briefly described. In
Section~\ref{sec:res}, our final results for the corrections considered
are presented for a sample physical value of $z = m_c^2/m_b^2$, namely
$z=0.04$. We conclude in Section~\ref{sec:summary}.

\newsection{Details of the calculation \label{sec:details}}

We work in the framework of an effective theory that is obtained from
the SM via the decoupling of the $W$ boson and all heavier
particles. The flavour-changing weak interaction terms that affect the
$b \to s \gamma$ transition take then the form\footnote{
Terms that contribute beyond the leading order in electroweak interactions 
and/or are suppressed by the small Cabibbo-Kobayashi-Maskawa 
matrix element $V_{ub}$ will be omitted here, as we focus on
${\mathcal O}(\al^2)$ effects.}
\be
{\mathcal L}_{\rm int} = \f{4 G_F}{\sqrt{2}} V_{ts}^* V_{tb} \sum_{i=1}^8 C_i(\mu_b) Q_i\, .
\ee
The $\overline{\rm MS}$-renormalized Wilson coefficients $C_i(\mu_b)$
are already known up to NNLO in QCD at the renormalization scale
$\mu_b \sim m_b$. Explicit expressions for the operators $Q_i$ can be
found, e.g., in Eq.~(1.6) of Ref.~\cite{Czakon:2015exa}.\footnote{
We shall strictly follow the notation of Ref.~\cite{Czakon:2015exa} throughout the current paper.}
For our present purpose, only three of them matter, namely
\mathindent0cm
\bea
Q_1   = (\bar{s}_L \gamma_{\mu} T^a c_L) (\bar{c}_L     \gamma^{\mu} T^a b_L),\hspace{11mm}
Q_2   = (\bar{s}_L \gamma_{\mu}     c_L) (\bar{c}_L     \gamma^{\mu}     b_L),\hspace{11mm}
Q_7  =  \f{e}{16\pi^2} m_b (\bar{s}_L \sigma^{\mu \nu}     b_R) F_{\mu \nu}\, .\hspace{-4cm}\nnb\\
\eea
\mathindent1cm

The weak radiative $b$-quark decay rate can be written as
\bea \label{rate}
\Gamma(b \to X_s^p \gamma) = 
\f{G_F^2 \alpha_{\mathrm em} m_{b,\rm pole}^5}{32 \pi^4} \left|V_{ts}^* V_{tb} \right|^2
\sum_{i,j=1}^8 C_i(\mu_b)\, C_j(\mu_b)\, \hat{G}_{ij}\, ,
\eea
where the quantities $\hat{G}_{ij}$ depend on the photon energy cut
$E_0$, the renormalization scale $\mu_b$, and the ratio $z =
m_c^2/m_b^2$ of the charm and bottom quark masses.\footnote{
The light $u$, $d$ and $s$ quark masses are set to zero in ${\mathcal
O}(\al^2)$ interference terms involving the operator $Q_7$ which we are
interested in here. However, they need to be retained in numerically
subleading terms to get rid of collinear divergences --- see, e.g.,
Refs.~\cite{Kaminski:2012eb,Asatrian:2013raa}.}
Their perturbative expansion in $\al$ reads
\be
\hat{G}_{ij} = \hat{G}^{(0)}_{ij} + \f{\al}{4\pi}\, \hat{G}^{(1)}_{ij}
+ \left( \f{\al}{4\pi} \right)^2 \hat{G}^{(2)}_{ij} + {\mathcal O}(\al^3)\, .
\ee

Currently, the dominant uncertainty in Eq.~(\ref{rate}) arises from the
$z$-dependence of $\hat{G}_{17}^{(2)}$ and $\hat{G}_{27}^{(2)}$. These
are the very quantities for which the interpolation mentioned in the
Introduction has been applied. In the following discussion, we refer only
to $\hat{G}_{27}^{(2)}$, for brevity. The calculation
of $\hat{G}_{17}^{(2)}$ has been performed alongside -- it differs by
colour factors only. We shall present both results in
Section~\ref{sec:res}.

Let us split the unrenormalized (bare) interference term
$\hat{G}_{27}^{(2)\rm bare}$ into contributions from two-, three- and four-particle
final states
\be \label{split1}
\hat{G}_{27}^{(2)\rm bare} = \hat{G}_{27}^{(2)2P} + \hat{G}_{27}^{(2)3P} + \hat{G}_{27}^{(2)4P}\, .
\ee
The two-particle contribution $\hat{G}_{27}^{(2)2P}$ can be further
split into a sum of two types of interference
\be \label{split2}
\hat{G}_{27}^{(2)2P} = \Delta_{30} \hat{G}_{27}^{(2)2P} + \Delta_{21} \hat{G}_{27}^{(2)2P}\, , 
\ee
where each $\Delta_{kn}$ picks the interference of a $k$-loop
amplitude with the insertion of $Q_2$, and an $n$-loop amplitude with
the insertion of $Q_7$. There are only two terms in Eq.~(\ref{split2})
because the one-loop $b \to s \gamma$ matrix element of $Q_2$ turns
out to vanish.

In the next section, we will present separate results for
$\Delta_{30} \hat{G}_{27}^{(2)2P}$ and $\Delta_{21}
\hat{G}_{27}^{(2)2P}$. The former can be calculated in two 
ways. One is to compute the three-loop $b \to s \gamma$ matrix
element of $Q_2$, and multiply its real part\footnote{
The imaginary part would matter only for the ${\mathcal O}(V_{ub})$
correction that we neglect at the ${\mathcal O}(\al^2)$ level. Such an
effect drops out after CP-averaging anyway.}
by the (real) tree-level matrix element of $Q_7$. Such an
``amplitude'' approach was applied in
Ref.~\cite{Fael:2023gau}. Partial results for $\Delta_{30}
\hat{G}_{27}^{(2)2P}$ in Ref.~\cite{Greub:2023msv} were also obtained using
the same method.

Here, we apply the cut propagator approach, as in
Ref.~\cite{Czakon:2015exa} in the $m_c=0$ case. It requires
essentially the same effort as the amplitude approach in the case of
two-particle final states, but is likely more convenient for
higher-multiplicity final states. Following the well-known
procedure~\cite{Anastasiou:2002yz}, we express the phase-space
integrals in terms of regular loop integrals but with cut 
propagators, using the identity
\be
-2\pi i \delta(p^2-m^2) = \f{1}{p^2-m^2+i\varepsilon} - \f{1}{p^2-m^2-i\varepsilon}\, . 
\ee
Once this is done, the usual Integration-By-Parts (IBP) algorithms can
be used to express the quantities in question in terms of Master
Integrals (MIs). The same algorithms are applied to derive Differential
Equations (DEs) for the MIs, which is essential for their efficient
evaluation.
\begin{figure}[t]
\begin{tabular}{ccc}
\includegraphics[scale=0.8]{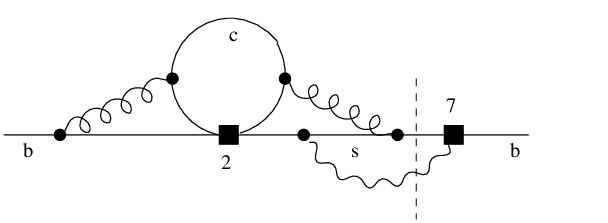} &&\\[-31.5mm]
&& \includegraphics[scale=0.8]{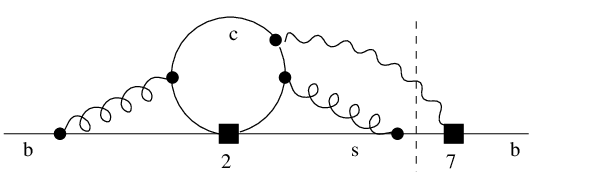}\\[6mm]
\includegraphics[scale=0.8]{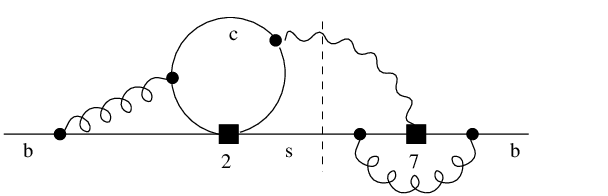} &&\\[-28mm]
&& \includegraphics[scale=0.8]{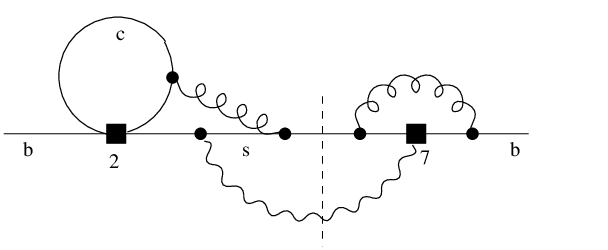}
\end{tabular}
\caption{ \sf Sample Feynman diagrams contributing to
$\Delta_{30} \hat{G}_{27}^{(2)2P}$ (upper row) and
$\Delta_{21} \hat{G}_{27}^{(2)2P}$ (lower row).
Black squares denote insertions of the $Q_2$ and $Q_7$ operators. The
vertical dashed lines indicate which propagators are cut.}
\label{fig:sample.diagrams}
\end{figure}

Sample four-loop propagator diagrams that contribute to
$\hat{G}_{27}^{(2)2P}$ in our approach are presented in
Fig.~\ref{fig:sample.diagrams}. Let us note that there exist physical
cuts in some of these diagrams that do not go through the photon line,
and therefore should not be included. This means that we neither take
the imaginary part of the whole four-loop propagator, nor take
advantage of the optical theorem.

We generate the necessary diagrams with the help of {\tt
QGRAF}~\cite{Nogueira:1991ex} and/or {\tt
FeynArts}~\cite{Kublbeck:1990xc,Hahn:2000kx} supplemented with
self-written codes. The Feynman--'t Hooft gauge fixing ($\xi=1$) is
used.  No diagrams with loop corrections on external (or cut) lines
are included, which matches the conventions of the corresponding
counterterm contributions in Eq.~(2.1) of
Ref.~\cite{Misiak:2017woa}. We skip the diagrams where no gluon
connects the charm loop (with the $Q_2$ vertex) to the rest of the
diagram. Such diagrams contain subdiagrams that either vanish or
sum up to zero. We also skip the diagrams with ghost or gluon one-loop
corrections on the gluon lines. Their effects are read out from the
corresponding massless quark loops, and included in our results in the
next section. Eventually, we are left with 198 four-loop propagator
diagrams that need to be calculated.
%

In each of the Feynman integrands, we average over the external $b$-quark
polarizations, and evaluate the necessary Dirac traces with the help
of {\tt FORM}~\cite{Ruijl:2017dtg}. Next, once all the Lorentz indices
have been contracted, the full $\hat{G}_{27}^{(2)2P}$ becomes a
linear combination of scalar integrals. Their reduction to MIs is
performed with the help of {\tt Kira}~\cite{Maierhofer:2017gsa,Klappert:2020nbg}
that generates and applies the IBP identities. Altogether, 447
MIs are found to be independent.

The calculation of MIs is performed with the help of {\tt
AMFlow}~\cite{Liu:2022chg}. An artificial imaginary mass-squared parameter
$\eta$ is introduced in each denominator, and the IBP method is used
to derive DEs in $\eta$ for the MIs. Next, the DEs are numerically
solved with initial conditions at very large $|\eta|$. All the
physical mass ratios and dimensionless kinematic invariants are
assigned fixed numerical values to facilitate the IBP reduction.

\newsection{Results \label{sec:res}}

In our case, the only physical mass ratio involved is $z =
m_c^2/m_b^2$. No kinematic invariants are present, as we deal
with a massive particle decaying to two massless particles.
To fix the numerical value of $z$, we follow the convention of
Ref.~\cite{Czakon:2015exa} where $z = m_c^{\overline{\rm
MS}}(\mu_c)/m_{b,{\rm kin}}$ was used in lower-order contributions,
where $m_{b,{\rm kin}} \simeq 4.564\,{\rm GeV}$ denotes the $b$-quark
mass in the kinetic scheme at $\mu_{\rm kin} = 1\,{\rm GeV}$. Below,
we present our final results for $z=0.04$ which corresponds to $\mu_c
\simeq m_b(m_b) \simeq 4.2\,{\rm GeV}$.

Our expressions for $\Delta_{30} \hat{G}^{(2)2P}_{27}$ and
$\Delta_{30} \hat{G}^{(2)2P}_{17}$ at $z=0.04$ read
\mathindent0cm
\bea
\Delta_{30} \hat{G}^{(2)2P}_{27}(z=0.04) &\simeq&
\f{0.181070}{\ep^3} -\f{6.063805}{\ep^2} -\f{34.087329}{\ep} -127.624515 \nnb\\
&+& \left(\f{0.482853}{\ep^2}+\f{4.093615}{\ep}+10.984004\right) n_b \nnb\\
&+& \left(\f{0.482853}{\ep^2}+\f{4.185427}{\ep}+19.194053\right) n_c \nnb\\
&+& \left(\f{0.482853}{\ep^2}+\f{4.135795}{\ep}+19.647238\right) n_l\; , \nnb\\[3mm]
\Delta_{30} \hat{G}^{(2)2P}_{17}(z=0.04) &\simeq&
-\f{1}{6}\Delta_{30} \hat{G}^{(2)2P}_{27}(z=0.04)
+\f{0.987654}{\ep^2} +\f{ 6.383643}{\ep} + 34.077780\, , \label{Delta30}
\eea
\mathindent1cm
where all the numerical coefficients have been truncated at the sixth decimal place.\\[-2mm]

As far as $\Delta_{21} \hat{G}^{(2)2P}_{27}$ and $\Delta_{21} \hat{G}^{(2)2P}_{17}$ are concerned, we find
\mathindent0cm
\bea
\Delta_{21} \hat{G}^{(2)2P}_{27}(z) &=& \f{368}{243\ep^3} + \f{736 - 324 f_0(z)}{243\ep^2} 
+ \f{1}{\ep} \left( \f{1472}{243} + \f{92}{729} \pi^2 - \f{8f_0(z)+4f_1(z)}{3} \right) + p(z)\; ,\nnb\\[3mm]
\Delta_{21} \hat{G}^{(2)2P}_{17}(z) &=& -\f{1}{6}\Delta_{21} \hat{G}^{(2)2P}_{27}(z)\, , \label{Delta21}
\eea
\mathindent1cm
where $p(z=0.04) \simeq 144.959811$. The large-$z$ expansion of $p(z)$ reads
\bea
p(z) &=& \f{138530}{6561}
- \f{3680}{729} \zeta(3)
- \f{6136}{243} L
+ \f{5744}{729} L^2
- \f{1808}{729} L^3\nnb\\[2mm]
&+& \f{1}{z} \left(
- \f{4222952}{1366875}
- \f{602852}{273375} L
+ \f{34568}{18225} L^2
- \f{532}{1215} L^3 \right)\nnb\\[2mm]
&+& \f{1}{z^2} \left(
- \f{33395725469}{26254935000} 
- \f{111861263}{93767625} L
+ \f{156358}{178605} L^2
- \f{172}{1215} L^3 \right) 
+ {\mathcal O}\left(\f{1}{z^3}\right)\, , 
%
%
%
\eea
with $L=\log z$. The NLO functions $f_0(z)$ and $f_1(z)$ are defined through
\be
\hat{G}_{27}^{(1)2P} = -\f{92}{81\ep} + f_0(z) + \epsilon f_1(z) + {\mathcal O}(\ep^2)\, . \label{def.f}
\ee
Their expansions around $z=0$ were originally found in
Refs.~\cite{Greub:1996jd} and~\cite{Misiak:2017woa},
respectively. Fully analytical expressions for them in terms of
harmonic polylogarithms have been recently determined in
Ref.~\cite{Fael:2023gau}. Their numerical values at $z=0.04$ are
$f_0(z=0.04) \simeq -6.371045$ and $f_1(z=0.04) \simeq -18.545805$.

Contributions from diagrams with quark loops on the gluon lines are
present in Eq.~(\ref{Delta30}) only. They are marked with $n_b=1$
(bottom loops), $n_c=1$ (charm loops), and $n_l=3$ (light quark
loops). The main new results of the current paper are the remaining
contributions that stem from diagrams with no quark loops on the gluon
lines.

In the case of Eq.~(\ref{Delta30}), we find perfect agreement
with the results of Ref.~\cite{Fael:2023gau}, after taking into
account their global normalization convention (see Eq.~(10)
there). To perform the comparison, we have relied on the
supplementary material to that paper where deep expansions around
$z=0$ of their quantities $t_2$ and $t_3$ are given. We can also
confirm the partial results of Ref.~\cite{Greub:2023msv}, once we
restrict to their subset of diagrams.

In the case of our Eq.~(\ref{Delta21}), analytical expressions for
all the $\f{1}{\ep^n}$ poles have been extracted from the former NLO
QCD calculations of the $\me{s\gamma|Q_2|b}_{\rm 2\;loop}$ and
$\me{s\gamma|Q_7|b}_{\rm 1\;loop}$~\cite{Asatrian:2006ph} 
matrix elements. They are in full agreement with our current
numerical results. As far as the finite contribution $p(z)$ is
concerned, it is determined for the first time here. It cannot be
extracted from the NLO results because dimensionally regulated
infrared (IR) divergences in $\me{s\gamma|Q_7|b}_{\rm 1\;loop}$
make the so-far-unknown higher-order terms in the $\ep$-expansion of
$\me{s\gamma|Q_2|b}_{\rm 2\;loop}$ relevant. The IR divergences will
cancel only after taking into account the yet
uncalculated NNLO contributions from diagrams with 3- and 4-body cuts.

\newsection{Summary and outlook \label{sec:summary}}

We evaluated the unrenormalized corrections $\hat{G}^{(2)2P}_{17}$ and
$\hat{G}^{(2)2P}_{27}$ to the perturbative $b \to s \gamma$ decay
rate. In the future, these need to be supplemented with
contributions from three- and four-body final states to get rid of IR
divergences. Then they can be renormalized using counterterm
contributions in Eq.~(2.1) of Ref.~\cite{Misiak:2017woa} where all the
necessary counterterm ingredients were calculated. We have already
tested such a renormalization in our expansions around the large-$m_c$
limit which provide the initial conditions required to
solve the DEs in $z$. The highest poles ($\f{1}{\ep^3}$ and
$\f{1}{\ep^2}$) are properly cancelled. However, a missing
piece with a $\f{1}{\ep}$ divergence in the three- and four-body bare
contributions was identified, and is currently being evaluated.

Given that the complete (partial) results for $\Delta_{30}
\hat{G}^{(2)2P}_{17}$ and $\Delta_{30} \hat{G}^{(2)2P}_{27}$ in
Ref.~\cite{Fael:2023gau} (\hspace{-0.0001mm}\cite{Greub:2023msv}) agree with our
findings, we have decided to present them now, and supplement with
$\Delta_{21} \hat{G}^{(2)2P}_{17}$ and $\Delta_{21}
\hat{G}^{(2)2P}_{27}$, even though no physical conclusion can be drawn
from unrenormalized results alone. Given the complexity of the
necessary calculations and the phenomenological relevance of the
expected ultimate result, presenting intermediate results allows for
valuable cross-checks and provides a boost for reaching the final
goal.

Our results in the previous section have been presented only for a
single value of $z$. In the case of $\Delta_{30} \hat{G}^{(2)2P}_{17}$
and $\Delta_{30} \hat{G}^{(2)2P}_{27}$, an exhaustive analysis of
$z$-dependence can be found in Ref.~\cite{Fael:2023gau}. As far as the
remaining contributions are concerned, we will study their
dependence on $z$ only at the level of the fully inclusive and
renormalized results for $\hat{G}^{(2)}_{17}$ and
$\hat{G}^{(2)}_{27}$.

\section*{Acknowledgements}

The research of M.~Czakon, T.~Huber, M.~Niggetiedt (partially)
and M.~Steinhauser was supported by the Deutsche
Forschungsgemeinschaft (DFG, German Research Foundation) under grant
396021762 --- TRR 257 ``Particle Physics Phenomenology after the Higgs
Discovery''. M.~Czaja and M.~Misiak acknowledge partial support by the
National Science Center, Poland, under the research project
2020/37/B/ST2/02746.  K.~Sch\"onwald was supported by the European
Research Council (ERC) under the European Union's Horizon 2020
research and innovation programme grant agreement 101019620 (ERC
Advanced Grant TOPUP).  M.~Niggetiedt was partially supported
by the Deutsche Forschungsgemeinschaft (DFG) under grant 400140256 -
GRK 2497: ``The physics of the heaviest particles at the Large Hardon
Collider.''  The work of A.~Rehman was supported by NSERC grant
SAPIN-2022-00020.


\begin{thebibliography}{99}
%
\bibitem{Chen:2001fja}
  S.~Chen \textit{et al.} (CLEO Collaboration),
  Phys.\ Rev.\ Lett.\ \textbf{87} (2001) 251807
  [hep-ex/0108032].
%
\bibitem{Saito:2014das}
  T.~Saito \textit{et al.} (Belle Collaboration),
  Phys.\ Rev.\ D \textbf{91} (2015) 052004
  [arXiv:1411.7198].
%
\bibitem{Belle:2009nth} 
  A.~Limosani \textit{et al.} (Belle Collaboration), 
  Phys. Rev. Lett. \textbf{103} (2009) 241801
  [arXiv:0907.1384].
%
\bibitem{Aubert:2007my}
  B.~Aubert \textit{et al.} (BaBar Collaboration),
  Phys.\ Rev.\ D \textbf{77} (2008) 051103
  [arXiv:0711.4889].
%
\bibitem{Lees:2012wg}
  J.~P.~Lees \textit{et al.} (BaBar Collaboration),
  Phys.\ Rev.\ D \textbf{86} (2012) 052012
  [arXiv:1207.2520].
%
\bibitem{Lees:2012ym}
  J.~P.~Lees \textit{et al.} (BaBar Collaboration),
  Phys.\ Rev.\ Lett.\  \textbf{109} (2012) 191801
  [arXiv:1207.2690].
%
\bibitem{Belle-II:2022hys}
  F.~Abudin\'en \textit{et al.} (Belle-II Collaboration),
  [arXiv:2210.10220].
%
\bibitem{HeavyFlavorAveragingGroup:2022wzx}
  Y.~S.~Amhis \textit{et al.} (Heavy Flavor Averaging Group (HFLAV)),
  Phys. Rev. D \textbf{107} (2023) 052008
  [arXiv:2206.07501].
%
\bibitem{ParticleDataGroup:2022pth}
  R.~L.~Workman \textit{et al.} (Particle Data Group),
  PTEP \textbf{2022} (2022) 083C01.
%
\bibitem{Bernlochner:2020jlt}
  F.~U.~Bernlochner \textit{et al.} (SIMBA Collaboration),
  Phys. Rev. Lett. \textbf{127} (2021) 102001
  [arXiv:2007.04320].
%
\bibitem{Dehnadi:2022prz}
  B.~Dehnadi, I.~Novikov and F.~J.~Tackmann,
  JHEP \textbf{2307} (2023) 214
  [arXiv:2211.07663].
%
\bibitem{Benzke:2010js}
  M.~Benzke, S.~J.~Lee, M.~Neubert and G.~Paz,
  JHEP \textbf{1008} (2010) 099
  [arXiv:1003.5012].
%
\bibitem{Gunawardana:2019gep}
  A.~Gunawardana and G.~Paz,
  JHEP \textbf{1911} (2019) 141
  [arXiv:1908.02812].
%
\bibitem{Benzke:2020htm}
  M.~Benzke and T.~Hurth,
  Phys. Rev. D \textbf{102} (2020) 114024
  [arXiv:2006.00624].
%
\bibitem{Benzke:2022cbw}
  M.~Benzke and T.~Hurth,
  [arXiv:2303.06447].
%
\bibitem{Hurth:2023paz}
  T.~Hurth and R.~Szafron,
  Nucl. Phys. B \textbf{991} (2023) 116200
  [arXiv:2301.01739].
%
\bibitem{Czakon:2015exa}
  M.~Czakon, P.~Fiedler, T.~Huber, M.~Misiak, T.~Schutzmeier and M.~Steinhauser,
  JHEP \textbf{1504} (2015) 168
  [arXiv:1503.01791].
%
\bibitem{Misiak:2010sk}
  M.~Misiak and M.~Steinhauser,
  Nucl.\ Phys.\ B \textbf{840} (2010) 271
  [arXiv:1005.1173].
%
\bibitem{Misiak:2015xwa}
  M.~Misiak, H.~Asatrian, R.~Boughezal, M.~Czakon, T.~Ewerth,
  A.~Ferroglia, P.~Fiedler, P.~Gambino, C.~Greub, U.~Haisch, T.~Huber,
  M.~Kami\'nski, G.~Ossola, M.~Poradzi\'nski, A.~Rehman, T.~Schutzmeier,
  M.~Steinhauser and J.~Virto,
  Phys.\ Rev.\ Lett.\  \textbf{114} (2015) 221801 
  [arXiv:1503.01789].
%
\bibitem{Misiak:2020vlo}
  M.~Misiak, A.~Rehman and M.~Steinhauser,
  JHEP \textbf{2006} (2020) 175
  [arXiv:2002.01548].
%
 \bibitem{Kou:2018nap}
  E.~Kou \textit{et al.} (Belle-II Collaboration),
  PTEP \textbf{2019} (2019) 123C01
  [arXiv:1808.10567].
%
 \bibitem{Fael:2023gau}
 M.~Fael, F.~Lange, K.~Sch\"onwald and M.~Steinhauser,
 JHEP \textbf{2311} (2023) 166
 [arXiv:2309.14706].
%
\bibitem{Greub:2023msv}
  C.~Greub, H.~M.~Asatrian, F.~Saturnino and C.~Wiegand,
  JHEP \textbf{2305} (2023) 201
  [arXiv:2303.01714].
%
\bibitem{Kaminski:2012eb}
  M.~Kami\'nski, M.~Misiak and M.~Poradzi\'nski,
  Phys.\ Rev.\ D \textbf{86} (2012) 094004
  [arXiv:1209.0965].
%
\bibitem{Asatrian:2013raa}
  H.~M.~Asatrian and C.~Greub,
  Phys.\ Rev.\ D \textbf{88} (2013) 074014
  [arXiv:1305.6464].
%
\bibitem{Anastasiou:2002yz}
  C.~Anastasiou and K.~Melnikov,
  Nucl.\ Phys.\ B \textbf{646} (2002) 220
  [hep-ph/0207004].
%
\bibitem{Nogueira:1991ex}
  P.~Nogueira,
  J.\ Comput.\ Phys.\ \textbf{105} (1993) 279.
%
\bibitem{Kublbeck:1990xc}
  J.~Kublbeck, M.~Bohm and A.~Denner,
  Comput.\ Phys.\ Commun.\  \textbf{60} (1990) 165.
%
\bibitem{Hahn:2000kx}
  T.~Hahn,
  Comput.\ Phys.\ Commun.\ \textbf{140} (2001) 418
  [hep-ph/0012260].
%
\bibitem{Misiak:2017woa}
  M.~Misiak, A.~Rehman and M.~Steinhauser,
  Phys.\ Lett.\ B \textbf{770} (2017) 431
  [arXiv:1702.07674].
%
\bibitem{Ruijl:2017dtg}
  B.~Ruijl, T.~Ueda and J.~Vermaseren,
  arXiv:1707.06453.
%
\bibitem{Maierhofer:2017gsa}
  P.~Maierh\"ofer, J.~Usovitsch and P.~Uwer,
  Comput. Phys. Commun. \textbf{230} (2018) 99,
  [arXiv:1705.05610].
%
\bibitem{Klappert:2020nbg}
  J.~Klappert, F.~Lange, P.~Maierh\"ofer and J.~Usovitsch,
  Comput. Phys. Commun. \textbf{266} (2021) 108024
  [arXiv:2008.06494].
%
\bibitem{Liu:2022chg}
 X.~Liu and Y.~Q.~Ma,
 Comput. Phys. Commun. \textbf{283} (2023) 108565
 [arXiv:2201.11669].
%
\bibitem{Greub:1996jd}
  C.~Greub, T.~Hurth and D.~Wyler,
  Phys.\ Lett.\ B \textbf{380} (1996) 385
  [hep-ph/9602281].
%
\bibitem{Asatrian:2006ph}
  H.~M.~Asatrian, A.~Hovhannisyan, V.~Poghosyan, T.~Ewerth, C.~Greub and T.~Hurth,
  Nucl. Phys. B \textbf{749} (2006) 325
  [hep-ph/0605009].
%
\end{thebibliography}
\end{document}